\documentclass[a4paper,11pt]{article}
\usepackage{pos}
\usepackage{wrapfig}

\title{The Compton Spectrometer and Imager Project for MeV Astronomy}
 \ShortTitle{The Compton Spectrometer and Imager}

\author*[a]{John A. Tomsick}

\affiliation[a]{Space Sciences Laboratory, 7 Gauss Way, University of California, Berkeley, CA 94720-7450, USA}

\forColl{COSI} 

\emailAdd{jtomsick@berkeley.edu}

\abstract{The Compton Spectrometer and Imager (COSI) is a 0.2--5\,MeV Compton telescope capable of imaging, spectroscopy, and polarimetry of astrophysical sources. Such capabilities are made possible by COSI's germanium cross-strip detectors, which provide high efficiency, high resolution spectroscopy and precise 3D positioning of photon interactions.  Science goals for COSI include studies of 0.511\,MeV emission from antimatter annihilation in the Galaxy, mapping radioactive elements from nucleosynthesis, determining emission mechanisms and source geometries with polarization, and detecting and localizing multimessenger sources.  The instantaneous field of view (FOV) for the germanium detectors is $>$25\% of the sky, and they are surrounded on the sides and bottom by active shields, providing background rejection as well as allowing for detection of gamma-ray bursts or other gamma-ray flares over $>$50\% of the sky.  We have completed a Phase A concept study to consider COSI as a Small Explorer (SMEX) satellite mission, and here we discuss the advances COSI-SMEX provides for astrophysics in the MeV bandpass.}

\FullConference{37$^{\rm{th}}$ International Cosmic Ray Conference (ICRC 2021)\\
		July 12th -- 23rd, 2021\\
		Online -- Berlin, Germany}

\begin{document}
\maketitle

\section{Overview}
\vspace{-0.2cm}
The Compton Spectrometer and Imager (COSI) is a wide-FOV telescope designed to survey the gamma-ray sky at 0.2--5\,MeV, performing high-resolution spectroscopy, imaging, and polarization measurements. COSI will map the Galactic positron annihilation emission, revealing the mysterious concentration of this emission near the Galactic center and the Galactic bulge in unprecedented detail.  COSI will elucidate the role of supernovae (SNe) and other stellar populations in the creation and evolution of the elements by mapping tracer elements. COSI will image $^{26}$Al with unprecedented sensitivity, perform the first mapping of $^{60}$Fe, search for young, hidden supernova remnants through $^{44}$Ti emission, and enable a host of other nuclear astrophysics studies. COSI will also study compact objects both in our Galaxy and AGN as well as gamma-ray bursts (GRBs), providing novel measurements of polarization as well as detailed spectra and light curves. The COSI FOV and localization capabilities make it powerful for searches of electromagnetic counterparts to gravitational wave and high-energy neutrino detections.  

\begin{wrapfigure}{r}{3.5in}
\centerline{\includegraphics[height=2.in,angle=0]{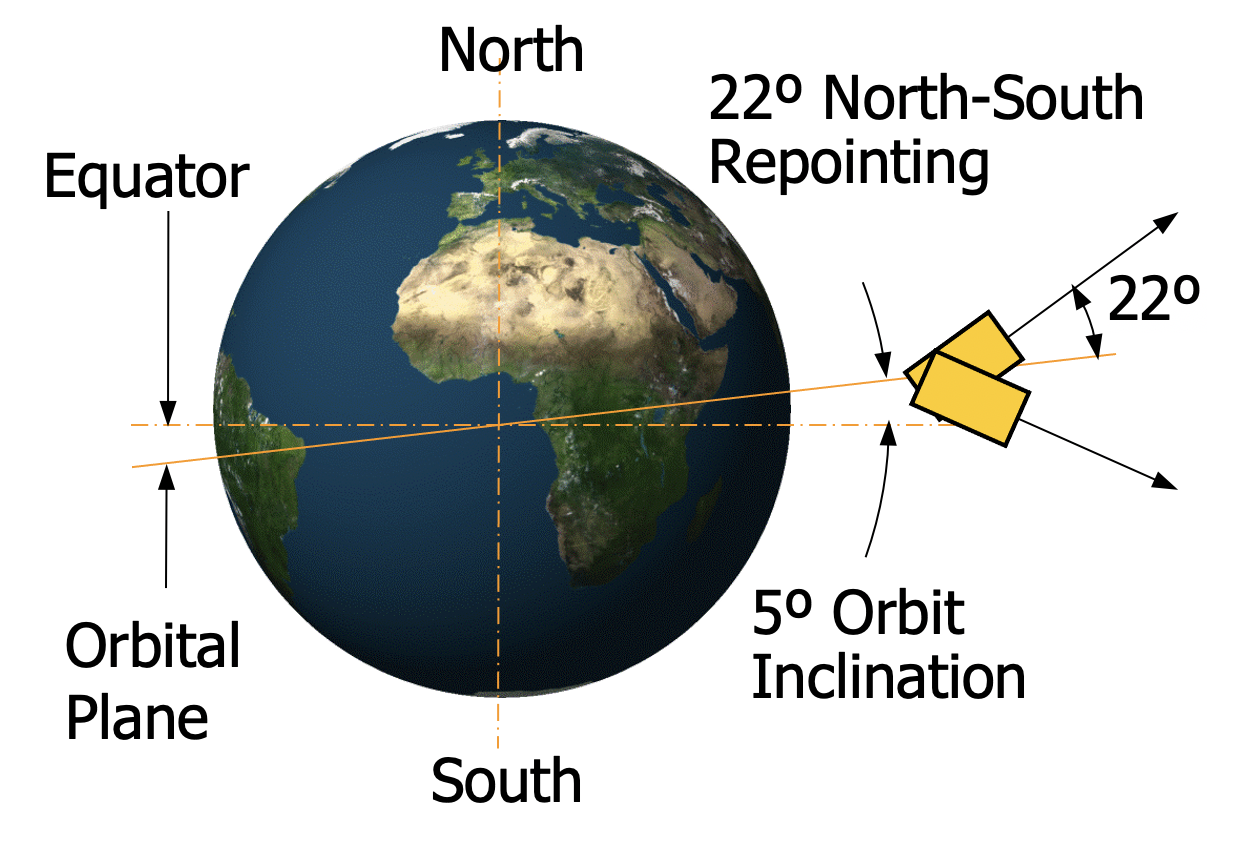}}
\caption{\small COSI (yellow rectangles) is shown in a low-inclination low-Earth orbit.  In survey mode, it monitors the entire sky on daily time scales via North-South repointing.  During the prime mission, COSI will spend $>$90\% of its time in survey mode.\label{fig:mission_overview}}
\end{wrapfigure}

The heart of COSI is a stacked array of germanium cross-strip detectors, which provide high efficiency, high resolution spectroscopy, and precise 3D positioning of photon interactions. As COSI is a Compton telescope, 3D positioning is required to carry out the Compton event reconstruction. We accomplish this using the versatile MEGAlib software package \citep{zoglauer06}.  The detectors are housed in a cryostat and cooled to $<$90\,K.  They are shielded on five sides, reducing the background and defining the FOV.  The COSI shields are active scintillators, extending COSI's FOV for detection of GRBs and other sources of gamma-ray flares.  COSI will be in low-Earth orbit and will spend most of the mission time in a survey mode, alternately pointing $22^{\circ}$ North of zenith to $22^{\circ}$ South of zenith every 12 hours to enable complete sky coverage over a day (see Figure~\ref{fig:mission_overview}).

\section{COSI Science Objectives}
\vspace{-0.25cm}
COSI has a large science portfolio that includes all-sky gamma-ray imaging for mapping electron-positron annihilation emission, radioactive decay lines, and Galactic diffuse emission. The long radioactive decay time scales of $^{26}$Al and $^{60}$Fe provide a look at where in the Galaxy nucleosynthesis has occurred over the past millions of years and where it is occurring now. While $^{60}$Fe is almost exclusively released into the interstellar medium (ISM) during core collapse supernovae (CCSNe), $^{26}$Al is also thought to be produced (and released) in the winds of massive stars.  $^{44}$Ti produced in SNe decays quickly by comparison, showing where recent supernova events have occurred in the last several hundred years and allowing for detailed studies of individual supernova 
\begin{wrapfigure}{r}{3.7in}
\centerline{\includegraphics[height=2.3in,angle=0]{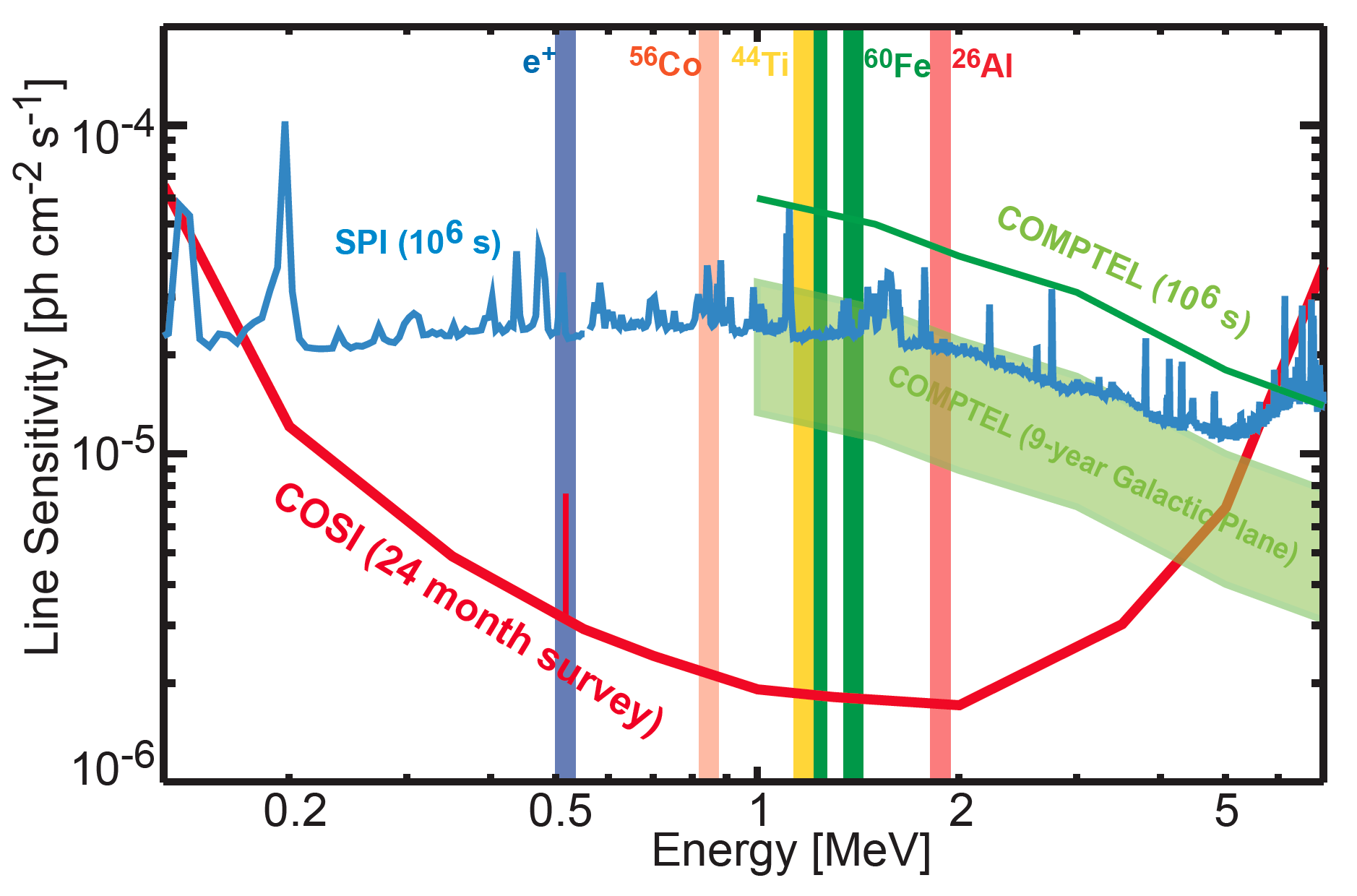}}
\caption{\small The COSI narrow line sensitivity for point sources (3$\sigma$) compared with COMPTEL and INTEGRAL/SPI.\label{fig:line_sensitivity}}
\end{wrapfigure}
remnants. The decay of $^{26}$Al, which produces a 1.809\,MeV gamma-ray and a positron, is the prominent known source of positrons through nucleosynthesis, but observations of the 0.511\,MeV positron annihilation line indicate that one or more additional unknown sources of positrons dominate their production. It is by comparing the full Galaxy maps of positron annihilation and radioactive isotopes from nucleosynthesis that we can understand the origin of the ``extra'' positrons. The high sensitivity that COSI provides for these emission lines compared to earlier measurements makes this goal possible (see Figure~\ref{fig:line_sensitivity}).

\subsection{Probing the Origin of Galactic Positrons}
\vspace{-0.25cm}
The production of positrons and their annihilation in the Galactic ISM is one of the pioneering topics of gamma-ray astronomy. Despite five decades of study since the initial detection of the 0.511\,MeV line from the inner Galaxy \citep{jhh72,lms78}, the origin of these positrons remains uncertain. INTEGRAL/SPI's imaging led to a map of the annihilation emission from the Galaxy that shows a bright bulge and fainter disk \citep{bouchet10}.  The angular resolution of the SPI map does not provide any definitive information about sub-structure in the emission.  No individual sources have been detected, and the origin of the bright bulge emission is unclear. The scale height of the disk emission measured by INTEGRAL/SPI is also uncertain, with Skinner et al. (\cite{skinner14}) finding an rms height of $\sim$$3^{\circ}$, while Siegert et al. (\cite{siegert16}) obtain $>$$9^{\circ}$.  This raises questions about the origin of the positrons as well as how far they propagate.  Even more fundamentally, the uncertain scale height means that the number of positrons in the disk is uncertain by a factor of more than three.  Thus, even though the 0.511\,MeV line is the strongest persistent gamma-ray signal, gaps remain in our understanding of this anti-matter component of our Milky Way \citep{kierans19}.

\subsection{Revealing Element Formation}
\vspace{-0.25cm}
The MeV bandpass includes nuclear emission lines that probe different physical processes in our Galaxy and beyond.  Long-lived isotopes such as $^{26}$Al (1.809\,MeV line) and $^{60}$Fe (1.173 and 1.333\,MeV lines), predominantly produced in SNe, provide information about the galaxy-wide star formation history, integrated over the past million years.  To first order, images of the Galaxy at these energies trace the last $\sim$10,000 CCSNe.  $^{44}$Ti (1.157\,MeV as well as 68 and 78\,keV lines in the hard X-ray range), with a half-life of 60 years, traces young Galactic SNe which occurred in the last few hundred years, and $^{56}$Co (0.847 and 1.238\,MeV) decays so rapidly (half-life: 77 days) that it is currently only seen by following up SNe in nearby galaxies. These lines allow us to trace the amount and distribution of these radioactive isotopes created in the underlying SNe giving us an independent and more direct view of Galactic chemical evolution \citep{timmes19}. 

\subsection{Insight into Extreme Environments with Polarization}
\vspace{-0.25cm}
Polarization measurements provide unique diagnostics for determining emission mechanisms and source geometries (e.g., magnetic field, accretion disk, and jet), but current results give just a glimpse into the potential for what we can learn.  In X-rays, the Crab nebula was detected at a level of 19\% \citep{novick72,weisskopf78}, but most sources show upper limits well below this, requiring a very sensitive X-ray polarimeter like the upcoming Imaging X-ray Polarimetry Explorer \citep{weisskopf16}.  Above 0.2\,MeV, in the COSI bandpass, much higher polarization levels have been measured for the Crab \citep{dean08,forot08} and for the accreting black hole Cygnus~X-1 \citep{laurent11,jourdain12}. 

A Compton telescope like COSI has intrinsic polarization sensitivity because the gamma-rays preferentially scatter in a direction perpendicular to their electric field.  Thus, with the sensitivity of COSI to polarization and the fact that COSI observes in a bandpass where the polarization level can be expected to be high for certain types of sources, COSI is likely to obtain polarization detections for many sources.  In addition to pulsars (like the Crab) and accreting stellar mass black holes (like Cyg~X-1), COSI will be able to measure polarization for several AGNs (like Cen~A), and its wide FOV allows for detection of large numbers of GRBs, for which polarization levels above 50\% have been reported \citep{mcconnell19}.  In a 2-year mission, COSI will see at least 40 GRBs bright enough to obtain a polarization measurement with a minimum detectable polarization of $<$50\%.

\subsection{Multimessenger Astrophysics}
\vspace{-0.25cm}
COSI significantly contributes to multimessenger astrophysics with its capability to find counterparts to multimessenger sources. This includes counterparts to gravitational wave events from binary neutron star (BNS) mergers as well as neutrino detections \citep{icecube18}.
 
For gravitational waves from BNS mergers like GW170817/GRB170817A \citep{abbott17b,goldstein17}, COSI is able to detect and localize the BNS by finding the associated short GRB.  Scintillator-based instruments have large FOVs but poor localization capabilities, and coded aperture mask instruments have good localization capabilities but smaller FOVs than COSI.  With Compton imaging, COSI combines the capability for sub-degree localizations with instantaneous coverage of $>$25\% of the sky to fulfill an important need for gamma-ray capabilities to detect and localize GRBs.  We estimate detection of $\sim$15-20 short GRBs in two years.  While the COSI shields will only provide rough localizations, they double the COSI FOV and allow comparison of the arrival time of short GRB to the arrival of the gravitational wave signal from BNS merger. While the cause of the 1.7\,sec delay of GRB170817A is uncertain, it may relate to the time for internal shocks to form in the jet, the shock breakout time, or the time for the merging NSs to collapse into a black hole.  

A COSI launch in 2025 would mean that it is observing when A+ generation gravitational wave detectors are planned to be in operation.  In the A+ era, the joint BNS-GRB detection range is 620\,Mpc (see Table 4 in \cite{burns20}) for the three LIGO interferometers, which takes a near face-on distribution averaged over the full-sky.  A distance of 620\,Mpc corresponds to $z=0.13$, and 14\% of short GRBs are within that redshift \citep{burns20}.  Thus, we would predict joint COSI and gravitational wave detections of 4.2-5.6 events during the 2 year COSI prime mission.

\section{COSI Instrument}
\begin{wrapfigure}{r}{3.0in}
\vspace{0cm}
\centerline{\includegraphics[height=2.2in,angle=0]{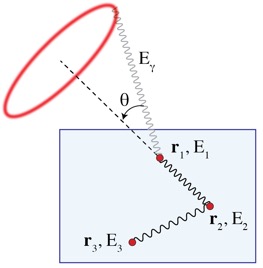}}
\vspace{0.0cm}
\caption{\small Schematic of interactions for a single gamma-ray in a Compton telescope.
\label{fig:schematic}}
\end{wrapfigure}

\vspace{-0.25cm}
The COSI instrument utilizes Compton imaging of gamma-ray photons \citep{vonballmoos89,bj00}.  An incoming photon of energy $E_{\gamma}$ undergoes a Compton scatter at a polar angle $\theta$ with respect to its initial direction at the position $r_{1}$, creating a recoil electron of energy $E_{1}$ that induces the signal measured in the detector (see Figure~\ref{fig:schematic}). The scattered photon then undergoes a series of one or more interactions, which are also measured. The Compton formula relates the initial photon direction to the scatter direction (measured direction from $r_{1}$ to $r_{2}$) and the energies of the incident and scattered gamma-rays. 

\begin{figure}[t]
\begin{center}
\includegraphics[height=3.0in,angle=0]{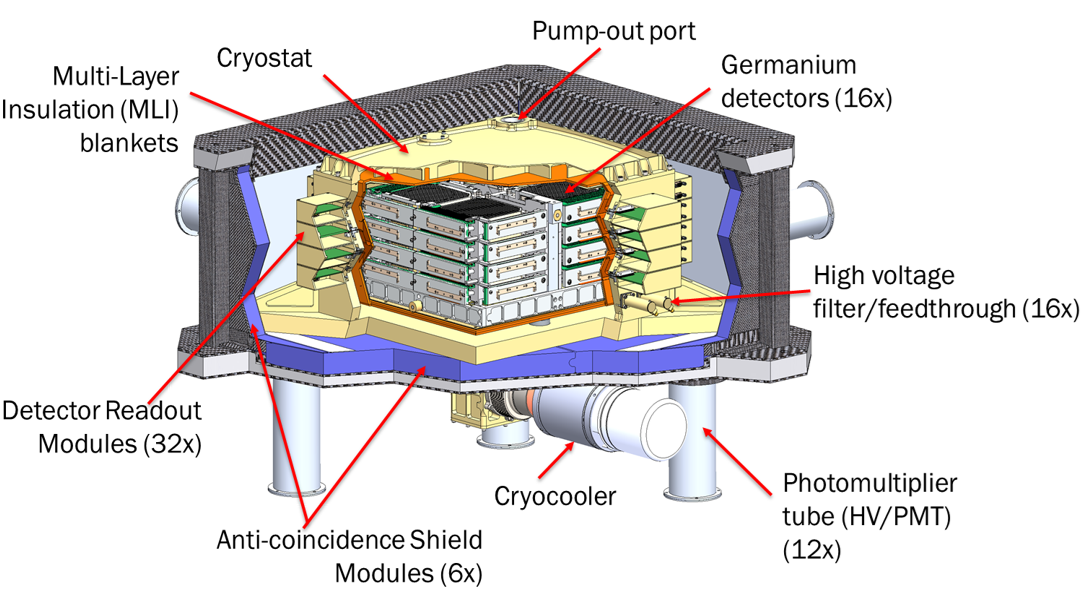}
\end{center}
\vspace{-0.5cm}
\caption{\small 
Cutaway view of the COSI instrument.  Each germanium detector is $8\times8\times1.5$\,cm$^{3}$.  The BGO shield box is $42\times46\times20$\,cm$^{3}$.
\label{fig:cosi_instrument}}
\end{figure}

COSI employs a novel Compton telescope design using a compact array of cross-strip germanium detectors (GeDs) to resolve individual gamma-ray interactions with high spectral and spatial resolution. No other technology thus far tested exceeds COSI's spectral resolution and polarization capability in this energy range. The COSI array of 16 GeDs is housed in a common vacuum cryostat cooled by a CryoTel CT mechanical cryocooler. The GeDs are read out by a custom ASIC, which is integrated into the full data acquisition system. An active BGO shield encloses the cryostat on the sides and bottom to veto events from outside the FOV.  The COSI instrument is shown in Figure~\ref{fig:cosi_instrument}.

We have developed the COSI instrument through NASA's Astrophysics Research and Analysis (APRA) program, including high-altitude balloon flights in 2005, 2009, 2014, and 2016, providing a very important proof of concept.  During a 46-day flight in 2016, the COSI imaging, spectroscopy, polarization, and real-time localization capabilities were demonstrated.  We were able to detect, localize, and report GRB 160530A in real time \citep{tomsick16_cosi}.  An analysis of the GRB 160530A data resulted in an upper limit on the polarization of $<$46\% \citep{lowell17a}, and new techniques were developed for polarization studies \citep{lowell17b} and for calibrating the instrument \citep{lowell17_phd,yang18}.  The detection of the 0.511\,MeV emission and information about the line shape and spatial extent is reported in Kierans et al. (\cite{kierans20}). We have also used calibration measurements to benchmark the detector effects engine \citep{sleator19}.

The data from the 2016 flight have also been vital for completing the software pipeline for the COSI imaging technique.  We have demonstrated the modeling of extended source distributions, such as the 0.511\,MeV emission from the Galactic bulge \citep{siegert20}, and also used the Crab nebula and pulsar as a point source for testing the technique \citep{zoglauer21}.  The software allows for model fitting of the image, which can be used for testing between physically-motivated spatial sky distributions and measuring fluxes, extents, and significances of the emission that is detected.

COSI provides major improvements in capabilities over previous MeV instruments.  Compared to COSI-APRA, the number of detectors is increased from 12 to 16, and they have 2 times better position resolution due to the strip pitch being decreased from 2\,mm to 1.16\,mm, which translates into a similar improvement in angular resolution.  Also, being above the atmosphere provides enormous improvements.  The simple removal of atmospheric attenuation improves sensitivity by 3-4 times, depending on energy, and being above the atmosphere reduces background.  Even compared to much larger current and previous satellite missions, COSI will provide an order of magnitude improvement in narrow line sensitivity (see Figure~\ref{fig:line_sensitivity}).

\section{COSI Requirements}
\vspace{-0.25cm}
Table~1 shows a sample of the requirements for COSI to achieve the science goals outlined above.  A key requirement for COSI is having excellent spectral resolution for making images of line emission over the entire Galaxy.  The spectral resolution is important for being able to make images in narrow energy bands around the lines of interest to minimize background.  In addition, spectral resolution is used to measure line profiles in order to study the width of the 0.511\,MeV line as well as the $^{44}$Ti line, which provides information about SN explosions.  The required sensitivities and angular resolution allow for COSI to distinguish between different physically-motivated Galactic bulge 0.511\,MeV emission models.  Also, these requirements provide the first images of Galactic $^{60}$Fe, a sensitive search for young $^{44}$Ti-emitting SN remnants, and the ability to isolate $^{26}$Al emission from individual massive star clusters.  While the large FOV requirement allows for the full Galaxy to be covered, the main driver for covering $>$25\% of the sky is to detect enough GRBs to achieve COSI's polarization and short GRB localization goals.  

\begin{table}[h]
\caption{COSI Requirements\label{tab:requirements}}
\begin{minipage}{\linewidth}
\begin{center}
\small
\begin{tabular}{lc|lc}
\hline \hline
{\bf Parameter} & {\bf Requirement} & {\bf Parameter} & {\bf Requirement}\\ \hline
Energy Range & 0.2-3\,MeV & FOV & 25\% sky\\ \hline
Spectral Resolution (0.511\,MeV) & 2.6 keV rms & Spectral Resolution (1.157\,MeV) & 3.9 keV rms\\ \hline
Angular Resolution (0.511\,MeV) & 3.8$^{\circ}$ FWHM & Angular Resolution (1.809\,MeV) & 2.0$^{\circ}$ FWHM\\ \hline
Line Sensitivity\footnote{3$\sigma$ narrow line sensitivity in 2-years of survey observations in units of photons\,cm$^{-2}$\,s$^{-1}$.}  (0.511\,MeV) & $1.0\times 10^{-5}$ & Line Sensitivity$^{a}$ (1.809\,MeV) & $3.0\times 10^{-6}$\\ \hline
Flux limit for polarization\footnote{Detection limit for 2-years of survey observations in units of erg\,cm$^{-2}$\,s$^{-1}$.} & $1.4\times 10^{-10}$ & GRB polarization fluence limit\footnote{For 50\% minimum detectable polarization.  The fluence limit is in units of erg\,cm$^{-2}$.} & $5.0\times 10^{-6}$\\ \hline
\end{tabular}
\end{center}
\end{minipage}
\end{table}

\section{Data Challenge Project}
\vspace{-0.25cm}
To improve the user-friendliness of the COSI analysis software and to prepare the high-energy community for analyzing COSI data, the COSI group is carrying out a Data Challenge project. For this project, the COSI team will publicly release a set of astrophysical source simulations, the full COSI imaging response, and documentation of our python-based analysis software (COSIpy).  Documentation and ``cookbook'' examples will be provided to the participants.  In addition, the COSI-APRA balloon data will be released to allow people to apply what they have learned to real data.

\section{Conclusions and Schedule}
\vspace{-0.25cm}
COSI will open up the MeV energy band to explore the science of positrons, nucleosynthesis, polarization studies of GRBs and accreting black holes, and multimessenger astrophysics.  With its large FOV and observing strategy, it will cover the entire sky in the 0.2-5 MeV bandpass every day.  The COSI team will be starting the Data Challenge in the latter part of 2021, and updates will be provided on the COSI website (http://cosi.ssl.berkeley.edu).  If COSI is selected for flight, it will enter Phase B in January 2022 and launch in late-2025.


\begin{thebibliography}{99}

\bibitem[\protect\astroncite{{Abbott} et~al.}{2017}]{abbott17b}
{Abbott}, B.~P., {Abbott}, R., {Abbott}, T.~D., et~al.\  2017, Physical Review
  Letters, 119, 161101

\bibitem[\protect\astroncite{{Boggs} \& {Jean}}{2000}]{bj00}
{Boggs}, S.~E., \& {Jean}, P.  2000, A\&AS, 145, 311

\bibitem[\protect\astroncite{{Bouchet} et~al.}{2010}]{bouchet10}
{Bouchet}, L., {Roques}, J.~P., \& {Jourdain}, E.  2010, ApJ, 720, 1772

\bibitem[\protect\astroncite{{Burns}}{2020}]{burns20}
{Burns}, E.,  2020, Living Reviews in Relativity, 23, 4

\bibitem[\protect\astroncite{{Dean} et~al.}{2008}]{dean08}
{Dean}, A.~J., {Clark}, D.~J., {Stephen}, J.~B., et~al.\  2008, Science, 321,
  1183

\bibitem[\protect\astroncite{{Forot} et~al.}{2008}]{forot08}
{Forot}, M., {Laurent}, P., {Grenier}, I.~A., {Gouiff{\`e}s}, C., \& {Lebrun},
  F.  2008, ApJ, 688, L29

\bibitem[\protect\astroncite{{Goldstein} et~al.}{2017}]{goldstein17}
{Goldstein}, A., {Veres}, P., {Burns}, E., et~al.\  2017, ApJ, 848, L14

\bibitem[\protect\astroncite{{IceCube Collaboration,~}
  et~al.}{2018}]{icecube18}
{IceCube Collaboration,~}{Aartsen}, M.~G., {Ackermann}, M., et~al.\  2018,
  Science, 361, 147

\bibitem[\protect\astroncite{{Johnson} et~al.}{1972}]{jhh72}
{Johnson}, W.~N., {Harnden}, F.~R., \& {Haymes}, R.~C.  1972, ApJ, 172, L1

\bibitem[\protect\astroncite{{Jourdain} et~al.}{2012}]{jourdain12}
{Jourdain}, E., {Roques}, J.~P., {Chauvin}, M., \& {Clark}, D.~J.  2012, ApJ,
  761, 27

\bibitem[\protect\astroncite{{Kierans} et~al.}{2019}]{kierans19}
{Kierans}, C., {Beacom}, J.~F., {Boggs}, S., et~al.\  2019,
\newblock in BAAS, Vol.~51,  256

\bibitem[\protect\astroncite{{Kierans} et~al.}{2020}]{kierans20}
{Kierans}, C.~A., {Boggs}, S.~E., {Zoglauer}, A., et~al.\  2020, ApJ, 895, 44

\bibitem[\protect\astroncite{{Laurent} et~al.}{2011}]{laurent11}
{Laurent}, P., {Rodriguez}, J., {Wilms}, J., et~al.\  2011, Science, 332, 438

\bibitem[\protect\astroncite{{Leventhal} et~al.}{1978}]{lms78}
{Leventhal}, M., {MacCallum}, C.~J., \& {Stang}, P.~D.  1978, ApJ, 225, L11

\bibitem[\protect\astroncite{{Lowell}}{2017}]{lowell17_phd}
{Lowell}, A.,  2017,
\newblock Ph.D. thesis, University of California, Berkeley

\bibitem[\protect\astroncite{{Lowell} et~al.}{2017a}]{lowell17a}
{Lowell}, A.~W., {Boggs}, S.~E., {Chiu}, C.~L., et~al.\  2017a, ApJ, 848, 120

\bibitem[\protect\astroncite{{Lowell} et~al.}{2017b}]{lowell17b}
{Lowell}, A.~W., {Boggs}, S.~E., {Chiu}, C.~L., et~al.\  2017b, ApJ, 848, 119

\bibitem[\protect\astroncite{{McConnell} et~al.}{2019}]{mcconnell19}
{McConnell}, M., {Ajello}, M., {Baring}, M., et~al.\  2019,
\newblock in BAAS, Vol.~51,  100

\bibitem[\protect\astroncite{{Novick} et~al.}{1972}]{novick72}
{Novick}, R., {Weisskopf}, M.~C., {Berthelsdorf}, R., {Linke}, R., \& {Wolff},
  R.~S.  1972, ApJ, 174, L1

\bibitem[\protect\astroncite{{Siegert} et~al.}{2020}]{siegert20}
{Siegert}, T., {Boggs}, S.~E., {Tomsick}, J.~A., et~al.\  2020, ApJ, 897, 45

\bibitem[\protect\astroncite{{Siegert} et~al.}{2016}]{siegert16}
{Siegert}, T., {Diehl}, R., {Khachatryan}, G., et~al.\  2016, A\&A, 586, A84

\bibitem[\protect\astroncite{{Skinner} et~al.}{2014}]{skinner14}
{Skinner}, G., {Diehl}, R., {Zhang}, X., {Bouchet}, L., \& {Jean}, P.  2014,
\newblock in Proceedings of ``10th INTEGRAL Workshop: A Synergistic View of the
  High-Energy Sky'', ~10

\bibitem[\protect\astroncite{{Sleator} et~al.}{2019}]{sleator19}
{Sleator}, C.~C., {Zoglauer}, A., {Lowell}, A.~W., et~al.\  2019, Nuclear
  Instruments and Methods in Physics Research A, 946, 162643

\bibitem[\protect\astroncite{{Timmes} et~al.}{2019}]{timmes19}
{Timmes}, F., {Fryer}, C., {Timmes}, F., et~al.\  2019,
\newblock in BAAS, Vol.~51, ~2

\bibitem[\protect\astroncite{{Tomsick}}{2016}]{tomsick16_cosi}
{Tomsick}, J.~A.,  2016, GRB Coordinates Network, 19473, 1

\bibitem[\protect\astroncite{{von Ballmoos} et~al.}{1989}]{vonballmoos89}
{von Ballmoos}, P., {Diehl}, R., \& {Schoenfelder}, V.  1989, A\&A, 221, 396

\bibitem[\protect\astroncite{{Weisskopf} et~al.}{2016}]{weisskopf16}
{Weisskopf}, M.~C., {Ramsey}, B., {O'Dell}, S., et~al.\  2016,
\newblock in Space Telescopes and Instrumentation 2016: Ultraviolet to Gamma
  Ray, Vol. 9905, SPIE Proceedings,  990517

\bibitem[\protect\astroncite{{Weisskopf} et~al.}{1978}]{weisskopf78}
{Weisskopf}, M.~C., {Silver}, E.~H., {Kestenbaum}, H.~L., {Long}, K.~S., \&
  {Novick}, R.  1978, ApJ, 220, L117

\bibitem[\protect\astroncite{{Yang} et~al.}{2018}]{yang18}
{Yang}, C.~Y., {Lowell}, A., {Zoglauer}, A., et~al.\  2018,
\newblock in Space Telescopes and Instrumentation 2018: Ultraviolet to Gamma
  Ray, Vol. 10699, Society of Photo-Optical Instrumentation Engineers (SPIE)
  Conference Series,  106992K

\bibitem[\protect\astroncite{{Zoglauer} et~al.}{2006}]{zoglauer06}
{Zoglauer}, A., {Andritschke}, R., \& {Schopper}, F.  2006, New Astronomy
  Reviews, 50, 629

\bibitem[\protect\astroncite{{Zoglauer} et~al.}{2021}]{zoglauer21}
{Zoglauer}, A., {Siegert}, T., {Lowell}, A., et~al.\  2021, arXiv e-prints,
  arXiv:2102.13158

\end{thebibliography}

\clearpage
\section*{Full Authors List: COSI Collaboration}
%
%
\scriptsize
\noindent
John A. Tomsick$^1$, 
Steven E. Boggs$^{1,2}$,
Andreas Zoglauer$^1$,
Eric Wulf$^3$,
Lee Mitchell$^3$,
Bernard Phlips$^3$,
Clio Sleator$^3$,
Terri Brandt$^4$,
Albert Shih$^4$,
Jarred Roberts$^{1,2}$,
Pierre Jean$^5$,
Peter von Ballmoos$^5$,
Juan Martinez Oliveros$^1$,
Alan Smale$^4$,
Carolyn Kierans$^4$,
Dieter Hartmann$^6$,
Mark Leising$^6$,
Marco Ajello$^6$,
Eric Burns$^7$,
Chris Fryer$^8$,
Pascal Saint-Hilaire$^1$,
Julien Malzac$^5$,
Fabrizio Tavecchio$^9$,
Valentina Fioretti$^9$,
Andrea Bulgarelli$^9$,
Giancarlo Ghirlanda$^9$,
Hsiang-Kuang Chang$^{10}$,
Tadayuki Takahashi$^{11}$,
Kazuhiro Nakazawa$^{12}$,
Shigeki Matsumoto$^{11}$,
Tom Melia$^{11}$,
Thomas Siegert$^{13}$,
Alexander Lowell$^1$,
Hadar Lazar$^1$,
Jacqueline Beechert$^1$,
Hannah Gulick$^1$\\

%
\noindent
$^1$Space Sciences Laboratory, 7 Gauss Way, University of California, Berkeley, CA 94720-7450, USA.
$^2$Center for Astrophysics and Space Sciences, UC San Diego, 9500 Gilman Drive, La Jolla CA 92093, USA.
$^3$U.S. Naval Research Laboratory, 4555 Overlook Ave., SW Washington, DC 20375, USA.
$^4$NASA Goddard Space Flight Center, 8800 Greenbelt Road, Greenbelt, MD 20771, USA.
$^5$Institut de Recherche en Astrophysique et Planétologie, 9, avenue du Colonel Roche BP 44346 31028 Toulouse Cedex 4, France
$^6$Clemson University, South Carolina, USA.
$^7$Louisiana State University, Louisiana, USA.
$^8$Los Alamos National Laboratory, New Mexico, USA.
$^9$Istituto Nazionale di Astrofisica, Italy.
$^{10}$Insitute of Astronomy, National Tsing Hua University, Taiwan.
$^{11}$Kavli IPMU, The University of Tokyo, Japan.
$^{12}$Nagoya Univeristy, Japan.
$^{13}$Institut für Theoretische Physik und Astrophysik, Universität Würzburg, Campus Hubland Nord, Emil-Fischer-Str. 31, 97074, Würzburg, Germany.

\end{document}